\documentclass[10pt,b5paper,twoside]{padeu}  
\usepackage{epsfig,natbib_padeu}             

\graphicspath{{figs/},{./}}

\newcommand{\Bp}{B_{\phi}}
\newcommand{\pa}{\partial}

\def\Rs{R_{\odot}}
\newcommand{\ov}{\overline}
\newcommand{\vc}[1]{\mbox{\bf #1}}

\newcommand{\citen}[1]{\citeauthor{#1} \citeyear{#1}}

\begin{document}

\setcounter{page}{5}

\title{Helical Magnetic Fields in Solar Active Regions: Theory vs. Observations}
\titlerunning{Helicity in solar active regions}
\author{K. Petrovay\inst{1}, 
	P. Chaterjee\inst{2},
	A. Choudhuri\inst{2}}
\authorrunning{K. Petrovay et~al.}
\institute{\inst{1} E\"otv\"os University, Department~of Astronomy, 
           H-1518 Budapest, Pf.~32, Hungary \\
	   \inst{2} Department of Physics, Indian Institute of Science, 
	      Bangalore 560012, India
}
\email{\inst{1}K.Petrovay@astro.elte.hu
}


\abstract{The mean value of the normalized current helicity  $\alpha_p=\vc
B\cdot(\nabla\times\vc B)/B^2$ in solar active regions is on the order of
$10^{-8}\,$m$^{-1}$, negative in the northern hemisphere, positive in the
southern hemisphere. Observations indicate that this helicity has a subsurface
origin. Possible mechanisms leading to a twist of this amplitude in magnetic
flux tubes include the solar dynamo, convective buffeting of rising flux tubes,
and the accretion of weak external poloidal flux by a rising toroidal flux
tube. After briefly reviewing the observational and theoretical constraints on
the origin of helicity, we present a recently developed detailed model for
poloidal flux accretion.
}

\maketitle


\section{Introduction}

\begin{figure}[t]
\centering
\epsfig{figure=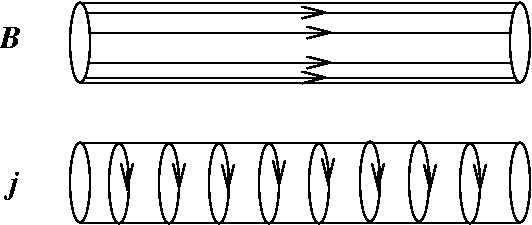, width=0.45\textwidth}\hfill
\epsfig{figure=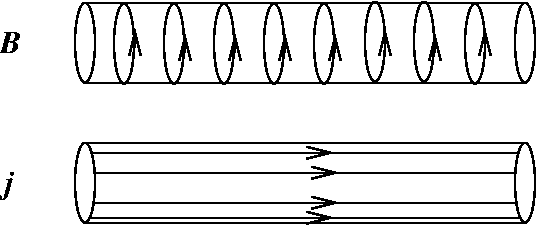, width=0.45\textwidth}\\
\bigskip
\epsfig{figure=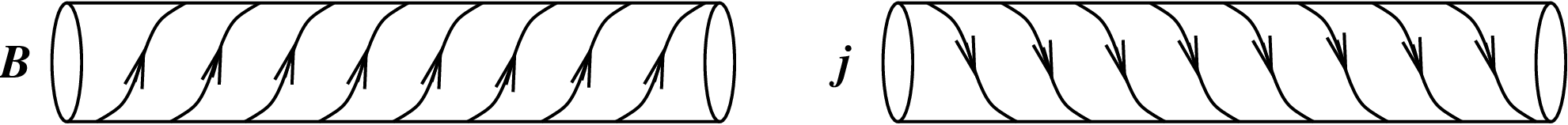, width=0.8\textwidth}
\caption{As Maxwell's equations are linear, the solutions can be superposed.
   The superposition of a straight flux tube (top left) and a 
   straight current tube (top right) results in a helical flux tube (bottom).}
   \label{fig:example1}
\end{figure}

Three-dimensional vector fields may possess {\it chirality} (handedness), i.e.
their structure may not be mirror symmetric. Concentrating on the magnetic
field $\vc B$, the simplest way to construct such a structure involves the
superposition of a magnetic flux tube and a  current tube. The result is a
twisted tube where both the magnetic field lines and the current lines have a
helical structure (Fig.1). This suggests that the helicity can be characterized
by the scalar product of $\vc B$ and $\vc j$; as the latter is proportional to
$\nabla\times\vc B$, the {\it current helicity} is defined as
\begin{equation} h_C=\vc B\cdot(\nabla\times\vc B) \end{equation}
Another quantity characterizing twist, often preferred for theoretical 
calculations is the {\it magnetic helicity}:
\begin{equation} h_M=\vc A\cdot\vc B\equiv\vc A\cdot(\nabla\times\vc A)    
\end{equation}
(To be precise, the above defined $h_C$ and $h_M$ are helicity densities,
helicity being their integral over a finite volume.)

Twist is just one form of helicity. As illustrated in Fig.~2, the twist of a
tube or band can be converted into {\it writhe} or {\it kink:}

\begin{figure}[h]
\centering
\epsfig{figure=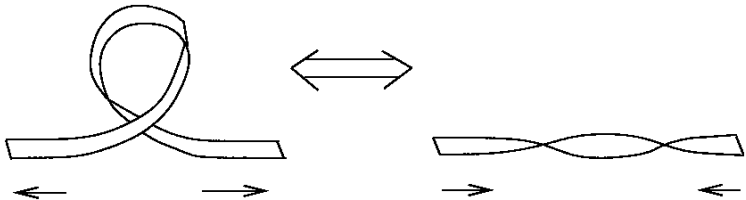, width=0.8\textwidth}
\caption{Twist and kink of a magnetic flux tube can be converted into each
other, as easily demonstrated by pulling the ends of a kinked paper tape, then
pushing them together again.}
   \label{fig:tape_kink}
\end{figure}

\section{Observations of helical structure on the Sun}

The apparent morphology of solar atmospheric structures, especially on
$H_\alpha$ images, is often vortex-like, twisted or helical. Early claims by
\citet{Hale:vortices} and \citet{Richardson:vortices} regarding the
predominance of one chirality or another on a given hemisphere were hard to
confirm owing to projection effects, a lack of information on 3D structure and
large statistical scatter. In any case, the existence of a hemispheric rule of
chirality was convincingly demonstrated at least in the case of quiescent
prominences and interplanetary magnetic clouds (\citen{Zirker+:chirality}).
Nevertheless, only direct magnetic measurements could provide convincing direct
evidence for chirality rules in active regions.

Most determinations of current helicity rely on a constant-$\alpha$ force-free
magnetic field model, constructed using the measured line-of-sight field
component as lower boundary condition. Such models are based on the assumption
that $\nabla\times\vc B=\alpha_p\vc B$ where $\alpha_p$ is constant. Multiplying
this relation by $\vc B$, it is clear that $\alpha_p=h_C/B^2$ is a normalized
measure of current helicity, with a dimension of 1/length. Its meaning is easy
to visualize, as $1/\alpha_p$ is the length along a flux tube over which he field
lines make a full circle around the tube.

Of the possible values of $\alpha_p$, a best fit is then chosen by comparing the
resulting magnetic field structure to either morphological features seen in
solar images taken in $H_\alpha$, EUV, etc. (\citen{Seehafer:helicity}) or,
more recently, to the horizontal field components as measured by vector
magnetographs (\citen{Pevtsov1995}). A third  method consists in simply taking
the ratio $\alpha_p=(\nabla\times\vc B)_n/B_n$, as measured by vector
magnetographs, and averaging it over the active region. Results derived by
different methods generally agree quite well
(\citen{Leka+Skumanich:twist.parameters}; \citen{Burnette+:twist.parameters}).

Currently all these methods yield a single mean value of $\alpha_p$ over the
whole active region. This is because the low S/N ratio of vector magnetograph
measurements inhibits a reliable study of current helicity distribution across
the plage; for the same reason, usually only pixels with relatively high field
strength ($B>100\,$G or so) are used in the determination. It is to be hoped
that future improvements in magnetograph sensibility will remedy this
situation.

These studies have now firmly established that active region magnetic fields
have helical structures, with a higher occurrence of negative helicity in the
northern hemisphere.  The observations show that the typical average value of
the current helicity parameter $\alpha_p$ in an active region is on the order
of $10^{-8}\,$m$^{-1}$ (\citen{vDG+:helicity.review}).

The tilt of the axis of bipolar active regions relative to the E-W direction is
generally interpreted as evidence for writhe in the emerging flux tubes giving
rise to active regions. The origin of this writhe is well understood: it is the
consequence of the Coriolis force acting on the downflows in the flux loop legs
(\citen{D'Silva+Choudhuri}). The sign of current helicity implied by the
observed tilts is positive on the northern hemisphere, i.e. opposite to the
twist measured in the solar photosphere; its amplitude, however, remains well
below the $\alpha_p$ values quoted above.

\section{The origin of helicity}

The basic question regarding the origin of the observed magnetic helicities is
whether they are generated after the emergence of the flux loop or the flux
emerges already in a helical form. Shearing of the photospheric footpoints of
magnetic loops due to differential rotation or smaller scale granular and
supergranular motions can, in principle, generate a helicity of right sign
(\citen{DeVore:diff.rot.helicity}); however, the amount of helicity generated
in this way seems to be insufficient to compensate for helicity losses in CME's
(\citen{vDG+:helicity.review}). In addition, in an important paper
\citet{Pevtsov+:subsurface.helicity} studied the time development of the
helicity in young, emerging active regions and found a good correlation between
emergence rate and helicity increase. In particular, further increase of
helicity ceases once the emergence of the loop, as measured by the increase in
footpoint separation and area growth, comes to a halt. This observation seems
to settle the issue in favor of a subsurface origin of the observed twist.

There are quite a few subsurface mechanisms which may naturally introduce twist
in the structure of emerging flux loops. Such proposed mechanisms include
helicity generation by the solar dynamo (\citen{Seehafer+:AdvSpRes}) and
buffeting of the rising flux tubes by helical turbulent motions 
(\citen{Longcope+:Sigma.effect}). A further possibility is the effect of 
Coriolis force on flows in rising flux loops (\citen{Fan+Gong:twist}). As we
mentioned above, this process is responsible for the generation of positive
writhe in active region flux tubes in the northern hemisphere, and thereby for
the observed tilt of active regions. As magnetic helicity is conserved in ideal
MHD (\citen{Berger:helicity.intro}), the same process should then also give
rise to a twist of opposite sense, so that the net helicity remains constant.
The amount of helicity generated by this process, however, is too low to
explain the observed values of $\alpha_p$. On the other hand, a strongly
twisted flux tube can develop a writhe by means of the kink instability. This
mechanism should lead to a positive correlation of twist and writhe, in
contrast to the helicity conservation argument above. Indeed, recent
observations by \citet{Lopez+:helicity} and \citet{Holder:helicity} indicate
that this mechanism may be important in the case of at least some active
regions with unusually high tilt.

\begin{figure}[ht]
\centering
\epsfig{figure=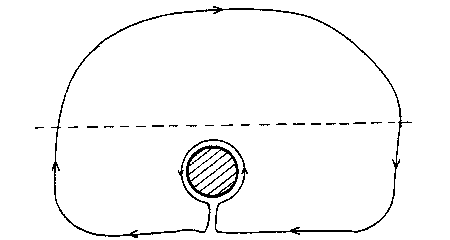, width=0.8\textwidth}
\caption{During the rise of a toriodal flux tube (here shown in cross section,
hatched) through the convective zone, field lines of the weak external poloidal
field may get wrapped around it. After \citet{Choudhuri:tube.dynamo}.}
   \label{fig:choudhuri_sketch}
\end{figure}

A further possible theoretical explanation of the observed helicity was
proposed by \citet{Choudhuri:tube.dynamo}, who suggested that the poloidal flux
in the solar convection zone (SCZ) gets wrapped around a rising flux tube, as
sketched in Fig.~3. \citet{Choudhuri+:helicity} later showed that this
mechanism gives rise to helicity of the same order as what is observed.
\citet{Choudhuri+:helicity} also used their dynamo model to calculate the
variation of helicity with latitude over a solar cycle and found that the
latitudinal distribution of helicity from their theoretical model is in broad
agreement  with observational data. 

If the magnetic flux in the rising flux tube is nearly
frozen, then we expect that the poloidal flux collected by it during its rise
through the SCZ would be confined in a narrow sheath at its outer periphery. In
order to produce a twist in the flux tube, the poloidal field needs to diffuse
from the sheath into the tube by turbulent diffusion. However, turbulent
diffusion is strongly suppressed by the magnetic field in the tube. This
nonlinear diffusion process was studied in an untwisted flux tube by 
\citet{Petrovay+FMI:erosion}. The model was subsequently successfully applied
for sunspot decay (\citen{Petrovay+vDG:decay1}). In a recent paper
(\citen{Chatterjee+:helicity.accretion}) we extended this model by including the
poloidal component of the magnetic field (i.e.\ the field which gets wrapped
around the flux tube) and we studied the evolution of the magnetic field in the
rising flux tube, as it keeps collecting more poloidal flux during its rise and
as turbulent diffusion keeps acting on it.

\begin{figure}[bt]
\centering
\includegraphics[width=0.7\textwidth]{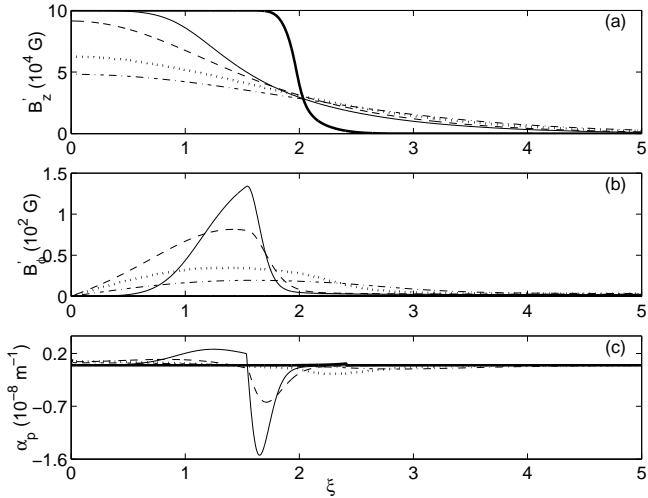}
\caption{Plots of $B_z'$, $\Bp'$ and  $\alpha_p$ as functions of $\xi$ for a
rising flux tube (case A). The different curves correspond
to the profiles of these quantities at the following positions of the flux
tube: $0.7 \Rs$ (thick solid), $0.85\Rs$ (solid), $0.9\Rs$ (dashed), $0.95\Rs$
(dotted), $0.98\Rs$ (dash-dotted).}
\label{fig:fig4}
\end{figure}

\begin{figure}[t]
\centering
\includegraphics[width=0.7\textwidth]{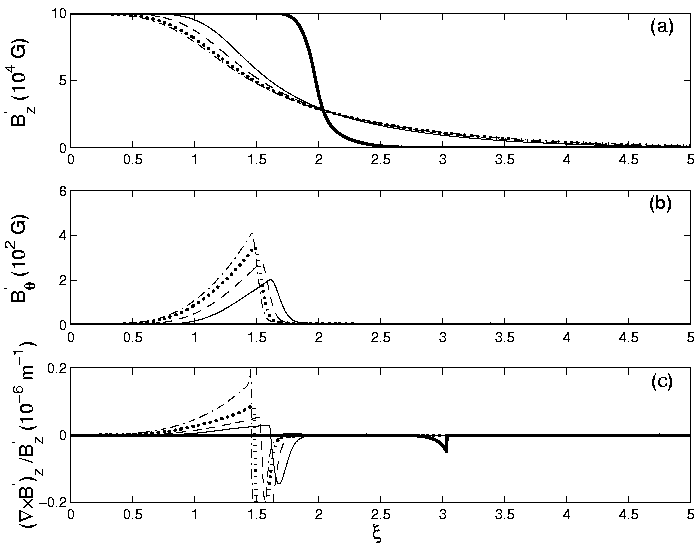} 
\caption{Same as Fig.~4 but the field  inside the flux tube is not allowed to
decrease below $3B_e$ at any height. (Case B)} 
\label{fig:fig5}
\end{figure}

Consider a straight, cylindrical, horizontal magnetic flux tube rising
through the solar convective zone. As all variables in this model depend only
on the radial distance $r$ from the tube axis and on time, we study the
wrapping of the large-scale poloidal field around the flux tube by considering
a radially symmetric accretion of azimuthal field by the flux tube. A further
complication is the expansion of the flux tube during its rise, due to the
decrease of the external pressure. This expansion is assumed to be
self-similar, so that the Lagrangian radial coordinate $\xi$ is related to the
Eulerian radius $r$ by $\xi = F (t) r $, where the expansion factor $F(t)$ was
taken from thin flux tube emergence models. In the Lagrangian frame, flux
density is rescaled as $ B'_z=B_z/ F^2 $ and $\Bp' =\Bp/F $. With these
notations and assumptions, the induction equation takes the form

\begin{equation}
\frac{\pa B'_z}{\pa t} 
= F^2 \frac{1}{\xi} \frac{\pa}{\pa \xi} 
\left( \eta \xi \frac{\pa B'_z}{\pa \xi} \right) 
\end{equation}
\begin{equation}
 \frac{\pa \Bp'}{\pa t} 
= F^2 \frac{\pa}{\pa \xi} \left[ \eta \frac{1}{\xi} 
\frac{\pa}{\pa \xi}(\xi \Bp') \right] - F \frac{\pa}{\pa \xi}(v \Bp').
\end{equation}
(Note that the advection term appears in the $\phi$ component only, as it does
not represent a truly radially symmetric inflow; instead, it is just designed to
mimick the effect of the wrapping of poloidal field lines around the rising
tube.)
Following \citet{Petrovay+FMI:erosion}, the magnetic diffusivity $\eta$ is
specified in the form
\begin{equation}
\eta = \frac{\eta_0}{1 + (B/B_{eq})^2}
\end{equation}
All our calculations are performed for a flux tube of flux $10^{22}\,$Mx and
initial field strength $10^5\,$G.

Numerical solutions of the equations are presented in Figures 4 and 5.  The
conclusions drawn from our model hinge on some assumptions, especially
concerning the subsurface magnetic field structure in the last phases of the
rise of the tube.  The field strength in the rising tube, as calculated from
thin flux tube emergence models, decreases well below the turbulent
equipartition value $B_e$ near the surface ($B_e^2/2\mu_0=\rho\ov{v^2}/2$). The
presence of 3000 G magnetic fields in sunspots is a compelling proof that
magnetic fields may never fall to such low values; in fact, at least in
photospheric layers, they are at about $3B_e$. This is presumably the result of
flux concentration processes such as turbulent pumping and convective collapse.
Thus, it may be more realistic not to allow the magnetic field to fall below
$3B_e$. Figures 4 and 5 present results without and with this constraint,
respectively.

Inspecting the lower panels of Figures~4 and 5 one finds that the typical value
of $\alpha_p$ in the internal parts of the flux tube is of order $\sim
10^{-8}\,$m$^{-1}$ at a depth of $0.85 \Rs$ in both cases.  However, as the
flux reaches the solar surface, in case A the $\Bp$ component spreads out due
to diffusion and its gradient becomes smaller, reducing $\alpha_p$ by about one
order of magnitude. Only if the magnetic field inside the flux tube remains
stronger than the equipartition field (the case B represented in  Figure~5) is
the $\Bp$ component unable to diffuse inside so that its gradient remains
strong and $\alpha_p$ is of order $\sim 10^{-8}\,$m$^{-1}$ even near the
surface.  This suggests that our case B may be closer to reality, i.e. during
the rise of the flux tube from $0.9 \Rs$ to $0.98 R_\odot$  effective flux
concentration processes are at work, keeping the field strength  at a value
somewhat above the equipartition level.

Note, however, that in the calculations presented here the amplitude and sign
of the poloidal field was assumed not to depend on depth. For alternative
assumptions, significantly different current helicities may result, so the
above conclusion should be treated with proper reservation. Details of the
radial dependence of the poloidal field strength may strongly depend on the
dynamo model.

A more robust feature of the current helicity distributions, present in all
the lower panels of our plots, is the presence of a ring around the tube with a
current helicity of the opposite sense. This is clearly the consequence of the
fact that on the outer side of the accreted sheath the radial gradient of the
azimuthal field, and thus the axial current, is negative. This is an inevitable
corollary of the present mechanism of producing twist in active regions. A
rather strong prediction of this model is, therefore, that a ring of reverse
current helicity should be observed on the periphery of active regions,
somewhere near the edge of the plage. 

\section{Conclusion}

One rather strong prediction of our model is the existence of a ring of reverse
current helicity on the periphery of active regions. On the other hand,  the
amplitude of the resulting twist (as measured by the mean current helicity in
the inner parts of the active region) depends sensitively on the assumed
structure (diffuse vs. concentrated/intermittent) of the active region magnetic
field right before its emergence, and on the assumed vertical profile of the
poloidal field. Nevertheless, a mean twist comparable to the observations can
result rather naturally in the model with the most plausible choice of
assumptions (case B). 

It is likely that the accretion of poloidal
fields during the rise of a flux tube is just one contribution to the
development of twist. Its importance may also be reduced by 3D effects:
considering the rise of a finite flux loop instead of an infinite horizontal
tube, the possibility exists for the poloidal field to ``open up'', giving way
to the rising loop with less flux being wrapped around it. It is left for later
multidimensional analyses of this problem to determine the importance of any
such reduction. In any case, the results presented above indicate that the
contribution of poloidal field accretion to the development of twist can be
quite significant, and under favourable circumstances it can potentially
account for most of the current helicity observed in active regions.


\begin{acknowledgement}
This research was carried out in the framework of the Indo-Hungarian 
Inter-Govern\-mental 
Science \& Technology Cooperation scheme, with the support
of the Hungarian Research \& Technology Innovation Fund and the Department of
Science and Technology of India. K.P. acknowledges support from the ESMN
network supported by the European Commission, and from the OTKA under grant
no.~T043741. P.C. acknowledges financial support from Council for Scientific
and Industrial Research, India under grant no.~9/SPM-20/2005-EMR-I.
\end{acknowledgement}



\end{document}